\documentstyle[preprint,prd,aps]{revtex}
\begin{document}
\draft
\preprint{RU9571, USC(NT)-95-3}
\title{An amplitude analysis of the $\overline{N}N \rightarrow \pi^- \pi^+$
reaction }
\author{W.M. Kloet}
\address{Department of Physics and Astronomy, Rutgers University, \\
Piscataway, NJ 08855, USA}
\author{F. Myhrer}
\address{Department of Physics and Astronomy, University of South Carolina, \\
Columbia, SC 29208, USA \\
and \\ NORDITA, Blegdamsvej 17, DK-2100 Copenhagen \O , Denmark }
\date{\today}
\maketitle
\begin{abstract}
A simple partial wave amplitude analysis of $\overline{p}p \rightarrow
\pi^-
\pi^+$
has been performed for data in the range p$_{\sl lab}$ = 360 -- 1000 MeV/c.
Remarkably few partial waves are required to fit the data,
while the number of required $J$ values barely changes over this
energy range.
However, the resulting set of partial wave amplitudes is not unique.
We discuss possible measurements with polarized beam and target
which will severely restrict and help resolve the present
analysis ambiguities.
New data from the reaction
$\overline{p}p \rightarrow \pi^0 \pi^0$
alone, are insufficient for that purpose.
\end{abstract}

\pacs{25.43.+t, 24.70.+s, 21.30.-x, 13.75.Cs, 13.75.Gx, 13.88.+e}

\narrowtext

\section{INTRODUCTION}

Although  the reactions $\overline{N}N \rightarrow \pi^-\pi^+$ and
$\overline{N}N
\rightarrow K^-K^+$
only account for less than one
percent of the $\overline{N}N$ total annihilation cross section,
they are two of the more basic annihilation
and subsequent hadronization reactions.
Therefore a careful study of these reactions can reveal details of the
underlying mechanisms and may clarify the nature of the degrees of
freedom necessary to describe these short range hadronic processes.
The new low energy data for $\overline{p}p \rightarrow \pi^- \pi^+$
(and $\overline{p}p \rightarrow K^- K^+$)
from LEAR~\cite{LEAR} compliment and extend the earlier
data~\cite{Eisen,Carteretal,Tanimori}
and show a rather rich energy and spin dependence.
The data below p$_{\sl lab}$~=~1.3 GeV/c
show considerable angular structure at each energy for both
the differential cross section and the analyzing power.
In addition, there
is considerable change
in the angular dependence of the observables
with increasing energy at these lower energies.
This is in contrast to the region above
p$_{\sl lab}$ = 1.3 GeV/c where there is considerably less
energy dependence in the angular structure
of these two observables.
Several analyses of the data of this reaction have been
performed~\cite{Carter,MartinP,MartinM,Oades,Bowcock} with the
pre-LEAR data.
In this paper we will concentrate on the reaction $\overline{p}p
\rightarrow \pi^- \pi^+$.

The recently published LEAR data were included in the newest
Durham analysis~\cite{Durham} which, like the older analysis,
writes the two
amplitudes (non-spinflip and spinflip) as one analytic
function of w = $e^{i \: \theta}$ where $\theta$ is the scattering angle.
This function is then parametrized by a
{\it finite} number of Barrelet zeros
in the complex w plane.
The Barrelet zeros {\it close} to the physical region show up as local
minima in the angular dependence of the spin dependent cross sections.
The invariant mass region of their analysis covers 1910 -- 2273 MeV,
and they require 8 -- 10 complex zeros to fit the data at a given energy.
The energy dependence of the resulting parametrization shows a
rich resonance structure in several partial waves.

The most recent analysis of this reaction by Hasan and Bugg~\cite{Bugg}
starts from the ansatz that each partial wave amplitude
of given $L$ and $J$
is a sum  of up to four resonances (of the same $J$) of different
mass and width. The maximum $J$ value is $J$ = 5, so there
are roughly one hundred parameters in their fit.
The many resonance parameters are then fitted
simultaneously to all available data
in the energy region of 1930 -- 2530 MeV.
Due to the above ansatz, the
resulting partial wave amplitudes exhibit counter-clockwise motion in
the Argand plots. However, they find only
clear peaks for the $J$ = 4 and 5 amplitudes.
In Ref.~\cite{Bugg} it is stated that the observables are approximately
symmetric about cos $\theta$ = 0.
As we see it, this statement does not reflect the data.
For example, the $\rm d\sigma/d\Omega$ is forward peaked at low energy
(p$_{\sl lab}$ = 360 MeV/c) and
backward peaked at energies above p$_{\sl lab}$~=~800 MeV/c
and no symmetry about cos $\theta$ = 0 is apparent.

\section{ANALYSIS}

In this paper we report on a different but very simple
amplitude analysis of the same data but we restrict our analysis to the
momentum range p$_{\sl lab}$ = 360 -- 988 MeV/c, the lowest measured momenta.
This p$_{\sl lab}$ range
corresponds to an invariant mass
region of 1910 -- 2078 MeV.
We reduce the theoretical input of this analysis to a minimum.
One main working hypothesis will be that very few partial waves
contribute to this particular annihilation reaction.
This hypothesis is based on the experience
gained by reproducing the observables
of this reaction with a simple black sphere
model at higher energies (p$_{\sl lab}$ $>$ 1 GeV/c)~\cite{TMK}.
In terms of the two independent helicity amplitudes
$\rm f_{++}$ and $\rm f_{+-}$
for this annihilation reaction,
the two measured observables are the differential cross section
\begin{equation}
d\sigma/d\Omega = (|f_{++}|^2 + |f_{+-}|^2)/2,
\end{equation}
and the analyzing power A$_{\sl on}$, defined by

\begin{equation}
\rm A_{\sl on} d\sigma/d\Omega = Im (f_{++} \: f_{+-}^* ) .
\end{equation}
We use here the convention that {\bf \^{n}} is the spin direction normal to
the scattering plane. The unit-vector {\bf \^{n}} is along
$\vec{p}$ x $\vec{q}$, where $\vec{p}$ is the antiproton
center-of-mass (CM) momentum and $\vec{q}$ is the CM momentum of $\pi^-$.
There are additional spin observables.
We define the longitudinal spin direction {\bf \^{l}}
along $\vec{p}$, and the transverse spin direction {\bf \^{s}} is
defined to be
along {\bf \^{n} $\times$ \^{l}}.
The analyzing powers A$_{\sl ol}$ and
A$_{\sl os}$ both vanish due to parity conservation.
Some  spin-correlation observables A$_{\sl ij}$,
for this reaction are however non-zero.
Of these observables
A$_{\sl nn}$ = 1, but A$_{\sl ll}$, A$_{\sl ss}$, A$_{\sl sl}$ and
A$_{\sl ls}$
are nontrivial,
with the identities A$_{\sl sl}$ = A$_{\sl ls}$  and A$_{\sl ss}$
=~--A$_{\sl ll}$.
The observables  A$_{\sl ss}$ and A$_{\sl ls}$ can be expressed in the
helicity amplitudes by respectively
\begin{equation}
\rm A_{\sl ss} \: d\sigma/d\Omega = (|f_{++}|^2 - |f_{+-}|^2)/2 ,
\end{equation}
and

\begin{equation}
\rm A_{\sl ls} \: d\sigma/d\Omega = {Re} (f_{++} \: f_{+-}^* ) .
\end{equation}
Unfortunately no data on spin correlations are as yet available.
As we shall discuss later, it is necessary to measure at least a third
observable of the reaction $\overline{p}p \rightarrow \pi^- \pi^+$ (and
$\overline{p}p \rightarrow K^- K^+$) to restrict
further the amplitude analysis.
The two helicity amplitudes are expanded in
$J \ne L$ spin-triplet partial wave amplitudes.
\begin{equation}
\rm f_{++} = \frac{1}{p} \sum_J \sqrt{J+\frac{1}{2}} \: ( \sqrt{J}\:
f_{J-1}^J    - \sqrt{J+1} \: f_{J+1}^J ) P_J(cos \theta)
\end{equation}
and

\begin{equation}
\rm f_{+-} = \frac{1}{p} \sum_J \sqrt{J+\frac{1}{2}} ( \frac{1}{\sqrt{J}} \:
 f_{J-1}^J  + \frac{1}{\sqrt{J+1}} \: f_{J+1}^J )
P_J^{\prime}(cos \theta) sin \theta ,
\end{equation}
where P$_J^{\prime}$
denotes the first derivative of the Legendre polynomial P$_J$.

In our analysis of the existing data of $\overline{p}p \rightarrow
\pi^- \pi^+$, we parametrize the partial wave amplitudes at each energy as

\begin{equation}
\rm  f_{\sl L}^{\sl J} = R_{\sl LJ} \: e^{\sl i \: \delta_{\sl LJ}}.
\end{equation}
where R$_{\sl LJ}$ and $\delta_{\sl LJ}$ are our free parameters.
At each energy we choose the maximum $J$ to be included in our
$\chi ^2$ search and we let the computer
search for the minimal $\chi ^2$ sum for a fit to both
$\rm d\sigma/d\Omega$ and A$_{\sl on}$.
In our fits we  choose
$\delta_{10}$ = 0  for the $^3P_0$ partial wave
whereas R$_{10}$ is a free parameter.
For all other $LJ$ values both phase and amplitude in Eq.(7)
are allowed to vary, to obtain the best fit.
When the complex amplitudes for the partial waves
$^3P_0$, $^3S_1$, $^3D_1$, etc. are determined by the minimal
$\chi^2$ search, we calculate as a test the
total reaction cross section with these amplitudes and compare with
the magnitude of the angular
integrated experimental cross section
\begin{equation}
\rm  \sigma = 2 \pi \int_{-1}^{1} d\sigma/d\Omega  \:  d ~ cos~ \theta
\end{equation}

Due in part to the incompleteness of the set of measured observables,
we do not find a unique solution, i.e., a unique
set of partial wave amplitudes in our
$\chi^2$ search.
The different solutions
depend  on the input-start values of the
amplitude parameters $R_{\sl LJ}$ and $\delta_{\sl LJ}$
and on the way the $\chi^2$ search is performed.
However, the minimal value of $\chi^2$ found in the various
possible fits are the same. If we had available data on
the reaction $\overline{p}p \rightarrow \pi^0 \pi^0 $ or on
other spin observables
it would be possible to restrict the choice among the various
amplitude-sets with equally good $\chi^2$.

The data for all measured energies
starting from p$_{\sl lab}$ = 360 MeV/c up to 1  GeV/c
can be fitted with partial wave amplitudes with total angular momentum
$J$ $\leq$ 3.
Once we have determined the amplitude values in Eq.(7) at one energy,
we use these as start values in our $\chi^2$ search at the neighboring
energies.
We have also fitted the data with a maximal $J$ = 4 using the same
procedure.
It appears that for $\overline{p}$ momenta, p$_{\sl  lab}$, above 988 MeV/c
the total $\chi^2$ does become somewhat better when we include
the $J$ = 4 partial wave amplitudes, but below  988 MeV/c
the improvement is marginal.
As examples we show two fits in Figs.~1 and 2 for p$_{\sl  lab}$ = 360 and
988 MeV/c.
By including the
$J$ = 4 amplitude, we introduce four more parameters, but the
$\chi^2$ per degree of freedom hardly improves.
It is surprising that so few partial waves with $J_{\sl  max}$ = 3
are sufficient in order to get
a satisfactory $\chi^2$  fit to the data.

At the same time we note that both $J$ = 2 and $J$ = 3
partial wave amplitudes are
essential even at the lowest measured energies
due to the shape of the asymmetry A$_{\sl  on}$.
The data for A$_{\sl  on}$ versus cos $\theta$
show two minima even at the lowest energy, p$_{\sl  lab}$ = 360 MeV/c,
and a local maximum close to cos $\theta$ = 0.
Previously~\cite{Bat} it has been noted that at least the
$J$=2 contribution
is needed to reproduce the shape of A$_{\sl  on}$.

In Table~I we show the $\chi^2$ per degree of freedom for
one set of partial wave amplitude parameters with $J_{\sl  max}$ = 3 as
well as the case
where $J_{\sl  max}$ =  2 or 4. Listed are the ten momenta
between p$_{\sl  lab}$ = 360 and 988 MeV/c, where there are available LEAR
data.
In Tables~II and III we give an example of a set of values for the
partial wave amplitudes R$_{\sl LJ}$ and their phases $\delta_{\sl LJ}$
found by our $\chi^2$ fit to the data.
The normalization of R$_{\sl LJ}$ is such that, if the momentum p in
Eqs.(5-6) is
expressed in GeV/c, the cross section defined in Eq.(1) is in $\mu$b/srad.
The corresponding $\chi^2$ values are those of Table~I for $J_{\sl  max}$ = 3.

In Figs.~3 and 4 we show the energy behaviour of our partial wave amplitude
parameters, R$_{\sl LJ}$ and $\delta_{\sl LJ}$ for all $LJ$ with
$J_{\sl  max}$ = 3.
The energy dependence indicates a growing importance of
$J$ = 2 and 3 with increasing energy for all solutions.
With respect to the $J$ = 0 and 1 amplitudes, we note that
the $^3D_1$ remains important through the entire
energy range while the  $^3P_0$ and $^3S_1$
amplitudes diminish
in importance with increasing energy for one set of solutions.
An interesting observation is
that for a given $J$ the L = J+1 contribution is at least as
important as the L = J--1 contribution. As discussed above
the $J$ = 2 and to a lesser extend $J$ = 3 amplitudes
already play a significant role at the lowest momentum 360 MeV/c.
The energy dependence of the phases is rather smooth.
Due to the ambiguities present in this analysis, we emphasize that the set of
parameters discussed here, and shown in Tables~II and III, is one
of many possible solutions that give a good fit to the data.

To indicate the level of the ambiguities in our
analysis, we give in Tables~IV and V a different set of
amplitude parameters.
This parameter set is a whole new family of amplitudes
and these amplitudes have their own
energy behaviour. It is remarkable that at all energies
the corresponding
values for the $\chi^2$ are practically identical to the $\chi^2$
tabulated in Table~I. This example bears out the
ambiguity of the present analysis, and illustrates our point that
all local $\chi^2$ minima of our different searches
lead to equally good fits to the data.
{\it Only new data at the same energies of other spin
observables like, for example,
A$_{\sl  ss}$ or A$_{\sl  ls}$,  can constrain these
ambiguities for $\overline{p}p \rightarrow \pi^- \pi^+$.}
Other analyses have used data of the reaction
$\overline{p}p \rightarrow \pi^0 \pi^0$,
which unfortunately will only put
constraints on the even $J$ (I = 0) amplitudes.

In Fig.~5 we show the total cross section for this reaction as a
function of p$_{\sl  lab}$  when we use the amplitudes found in the
$\chi^2$ search.
For both the parameter set of Tables~II and III, and
the alternative set of Tables~IV and V, the total cross section for
$\overline{p}p
\rightarrow \pi^- \pi^+$ as a function of p$_{\sl  lab}$ agrees with the
total cross section of Hasan {\it et al.} within a few percent.
The momentum or energy dependence of each
individual J-value contribution to the total cross section
is shown in Fig.~5 for the solution corresponding to Tables~II
and III. The alternative parameter set gives, as expected,
a different distribution of individual J-value contributions
but leads to the same total cross section.
For both parameter sets there is a considerable
contribution from $J$ = 2 at the lowest momentum,
p$_{\sl  lab}$ = 360 MeV/c. The $J$ = 3 contribution to the total
cross section,  is small at 360 MeV/c,
but has become substantial at 500 MeV/c.

Finally, in Fig.~6 we show examples of Argand plots for the amplitudes
$^3P_2, ^3F_2, ^3D_3, ^3G_3$.
The amplitudes in this figure are taken from Tables~II and III.
One should bear in mind that the largest $J$ value equals 3 in
the fit producing these
amplitudes. In addition, we are considering a very limited energy
range. Therefore, we cannot draw any firm conclusions about evidence
for resonances in any of these sets of amplitudes.

\section{Discussion of results}

It is clear from all our fits
that  at all momenta below 1 GeV/c very few partial waves suffice
to fit the data. Partial wave amplitudes
with $J$ = 0,1,2, and 3 are adequate in all cases. Adding the $J$ = 4
partial wave amplitude does
not improve the $\chi^2$ per degree of freedom in our analysis,
except for momenta close to 1 GeV/c.
This statement can be made, while
no theoretical bias  as to the energy- and analytical- behaviour of
the amplitudes has been imposed on our fits.
These results are not inconsistent
with the simple model analysis at the higher
energies~\cite{TMK}.
This earlier work~\cite{TMK} used a diffractive
scattering model from a simple black
or grey sphere which
could explain most of the features of the higher energy data (p$_{\sl  lab}$
above 1.5 GeV/c, and above 1.0 GeV/c for $\overline{p}p \rightarrow K^- K^+$).
In that model description of the data, the spin dependent forces
were assumed to act in the surface region only.
The idea was that since the central region was ``black'' no detailed
information
would escape from the central interaction region.
Only the more transparent surface region would
provide the spin-forces giving the asymmetries of this annihilation  reaction.
No apparent resonances were needed in that simple model description.

     In our analysis the data for A$_{\sl  on}$ with an apparent local
maximum close to cos $\theta$ = 0, and the
presence of two minima
even at the three lowest energies, requires the presence
of the $J$ = 2 and $J$ = 3 amplitudes.
The minimal
$\chi^2$ value with $J_{\sl  max}$ = 2 is much larger than for any of the
$J_{\sl  max}$ = 3 searches. Therefore, we conclude that both the
$J$ = 2 and $J$ = 3 amplitudes are necessary.
As mentioned before, adding $J$ = 4 does not improve the $\chi^2$
per degree of freedom.

For A$_{\sl  on}$ the experimental errors  are very large
at backward angles for the lowest energies and
this prevents us from distinguishing between the solutions that
include the $J$ = 4 amplitudes from  those that only include the
$J$ $\leq$ 3 amplitudes.
Note that at the lowest energies the $J$ = 4 amplitudes tend to
reduce the backward local maximum of A$_{\sl  on}$
compared to the $J_{\sl  max}$ = 3 analysis as indicated in Figs.~1 and 2.
More accurate data are needed
to settle the issue of the necessity of $J$ = 4 amplitudes at these energies.
{}From a pure $\chi^2$ point of view the $J$ = 4 amplitudes are not really
needed
for energies below 900 MeV/c.

The observed forward peaking in $\rm  d\sigma/d\Omega$
at 360 MeV/c, occurs because of a strong
cancellation between the even and odd $J$ amplitudes for backward angles
($\theta \approx 180^0$).
At our highest energies a backward peak
develops, due in part to a constructive interference for backward
scattering angles.
One notes that at the two highest energies
in this analysis (p$_{\sl lab}$~=~886 and 988 MeV/c), the forward peak
in $\rm d\sigma/d\Omega$ is much smaller in magnitude than the
backward peak.

{}From the amplitude set of Table~II (and Table~III) it can be seen
that the $^3P_0$ amplitude becomes less important with increasing energy
whereas for the alternative amplitude set of Tables~IV and V
the $^3P_0$ amplitude stays important.
In addition, observe that at the highest energies of this
analysis the amplitudes $R_{J+1,J}$ are often larger than the
corresponding $R_{J-1,J}$.
We looked for indications that the initial
$\overline{N}N$ tensor coupled partial waves would be tensor-eigenstates
as proposed by Dover and Richard~\cite{DR78}.
This can be checked by looking at the phase
correlation of $f^J_{J-1}$ and $f^J_{J+1}$, Eq.(7).
{}From the Tables~II through V we only find a phase difference
$|\delta^J_{J-1} - \delta^J_{J+1}|$
of 0$^{\circ}$ or 180$^{\circ}$ at very few energies.
Therefore, in general the initial $\overline{N}N$ states are not pure tensor
eigenstates and the  $f^J_{J-1}$ and $f^J_{J+1}$ are not correlated in
phase.
We should mention
that adding the two $J$ = 4 amplitudes of course changes the phases and
amplitudes of the lower $J$ amplitudes found by
imposing $J_{\sl max}$ = 3 in our $\chi^2$ search.
The reason is that the $J$ = 4 amplitudes can mimic part of the
behaviour of the lower partial wave amplitudes.

     At p$_{\sl lab}$ = 783 MeV/c we tested several aspects of our analysis
and also examined the hypothesis of Bowcock and Hodgson~\cite{Bowcock}
that only a few partial waves on the periphery of the interaction
region are active, and that the central partial wave amplitudes are
very small.
For this purpose we have made a $\chi^2$ search excluding the
$J$ = 0 amplitude and
including the $J$ = 4 amplitudes. In essence we use the
partial wave amplitudes $J$ = 1,2,3,4 in our fit.
This part of the analysis, inspired by the Bowcock and Hodgson
conjecture, is summarized in Fig.~7
where fits for the three cases $J$ = 0,1,2,3; $J$ = 0,1,2,3,4;
and $J$ = 1,2,3,4 at p$_{\sl lab}$~=~783 MeV/c are shown on the same graph.
All three cases of $J$-value combinations have comparable
total $\chi ^2$ per degree of freedom, which are respectively 1.49,
1.47, and 1.50.
Of course, the standard case of $J_{\sl max}$ = 3 has the
smallest number of parameters because the $J$ = 0 amplitude involves
only two parameters instead of the usual four for the $J$ $\neq$ 0
amplitudes.
This phenomenon does not extend to other combinations of $J$ values
(e.g. excluding both $J$~= 0 and $J$ = 1 partial waves), and we find
no clear evidence for the hypothesis that only the most peripheral
partial waves are active.
In fact with only three
partial wave amplitudes non-zero like $J$ = 1,2,3 or $J$ = 2,3,4,
the total $\chi^2$ per degree of freedom is far worse than what
is shown in Table~I for $J_{\sl max}$ = 3.
This peculiar phenomenon holds also at other energies.

\section{CONCLUSION}

     The fact that very few partial waves are needed in the analysis,
indicates that
this annihilation reaction is a very central process.
In contrast, to describe the
elastic $\overline{p}p \rightarrow \overline{p}p$ scattering at the same
energies requires
the contributions from partial waves of $J$ = 9 (and higher)
due to the long range pion exchange potential.

As opposed to Oakden and Pennington~\cite{Durham}, we
find no compelling evidence for resonance behaviour of our partial
wave amplitudes when plotted in an Argand diagram in the energy
range considered, see Figs.~3 and 6.
The approach of Oakden and Pennington has some similarities with ours
in the sense that both methods do not
assume a specific energy dependence of the
amplitudes and both make a sharp truncation of the partial wave series.
A different approach is taken by Hasan and Bugg~\cite{Bugg}
who by starting from a description in terms of a tower of resonances
assume a specific energy dependence.
A further, unbiased, more refined analysis of these very good data,
with the additional constraints of a third measured observable,
is needed to settle
the question of possible resonance behaviour of the amplitudes.

A possible theoretical constraint on the amplitudes behaviour
could come from the analytic continuation of the $\pi N \rightarrow
\pi N$ scattering amplitudes as is done by
H\"{o}hler and Pietarinen~\cite{Hoh}.
So far this type of analytic continuation from $\pi N \rightarrow \pi
N$ elastic
scattering amplitudes to $\overline{p}p \rightarrow \pi^- \pi^+$ amplitudes
is very involved and
has only been performed for $\overline{p}p$ sub-threshold
energies~\cite{Brown}.
See however the analysis of Martin and Oades~\cite{Oades} who did make
use of the crossed channel $\pi N$ elastic scattering data.
We
believe that further experimental investigation of the reaction
$\overline{p}p \rightarrow \pi^- \pi^+$
is a more straight forward alternative.

In summary, the present experimental data and several recent analyses
of these
data for both $\overline{p}p \rightarrow \pi^- \pi^+$ and $\overline{p}p
\rightarrow K^- K^+$ reactions
promises a better understanding of these simple,
but fundamental annihilation reactions.
The analysis described in this paper
suggests the following future experiments necessary in
order to clarify the understanding of
the $\overline{p}p \rightarrow \pi^-\pi^+$ annihilation reaction:

(i) First, measurements of this annihilation
reaction should be made at antiproton
momenta closer to threshold. At very low energies
even fewer partial wave amplitudes
would contribute and the ambiguities of an analysis
would be reduced. However, ambiguities would remain
at the energies we considered.

(ii) Second, data in the exact same momentum range for the reaction
$\overline{p}p \rightarrow \pi^0 \pi^0$,
which is described by the isoscalar
amplitudes $^3P_{0}, ^3P_{2},^3F_{2},^3F_{4},^3H_{4}$, etc.,
would certainly be extremely useful.
They would impose further constraints on the analysis of $\overline{p}p
\rightarrow \pi^- \pi^+$.
However, from the point of view of amplitude analysis, each helicity
amplitude would have then an I=0 and I=1 part, and
only the I=0 (J even) amplitudes are restricted by this reaction.

(iii) Third, a different and more direct constraint
on the analysis of this reaction would come
from data on the other spin observables for $\overline{p}p \rightarrow
\pi^- \pi^+$
with longitudinal and/or transverse polarized beam and target,
for example data on A$_{\sl ss}$ or A$_{\sl ls}$.
To illustrate the sensitivity of these two observables
we have given in Fig.~8
the predictions for A$_{\sl ls}$ and A$_{\sl ss}$ for the two parameter sets
of Tables~II, III and Tables~IV, V.
As is obvious from the figures the two predictions differ significantly
in the forward hemisphere where we believe experiments can be performed more
accurately.
Some spin correlation observables have
already been measured for
$\overline{p}p$ elastic and charge-exchange scattering at LEAR, and we
propose here the use of polarized antiprotons to obtain
data on  the observables A$_{\sl ls}$ or A$_{\sl ss}$ for $\overline{p}p
\rightarrow \pi^- \pi^+$.
Such an experiment would settle the debate about the possible resonances
inferred from analysis of this reaction.

\section{ACKNOWLEDGEMENTS}

We thank F. Bradamante for providing us with the LEAR data,
and R. Timmermans
for many useful discussions on the art of phase shift analyses. This work
is supported in part by NSF grants no. PHYS-9310124 and no. PHYS-9504866.

\begin{figure}
\caption[]{$\rm d\sigma/d\Omega$ and A$_{\sl on}$ at p$_{\sl lab}$ = 360 MeV/c.
The solid curves give the fit for $J_{\sl max}$ = 3 and the dashed curves
are the fit for $J_{\sl max}$ = 4.}
\end{figure}

\begin{figure}
\caption[]{$\rm d\sigma/d\Omega$ and A$_{\sl on}$ at p$_{\sl lab}$ =
988 MeV/c.
The solid curves give the fit for $J_{\sl max}$ = 3 and the dashed curves
are the fit for $J_{\sl max}$ = 4.}
\end{figure}

\begin{figure}
\caption[]{The amplitudes R$_{\sl LJ}$ of our fit ($J_{\sl max}$ = 3) versus
p$_{\sl lab}$. The curves are only meant to guide the eye.
The upper solid curve is $J$ = 0, dashed curves are $J$ = 1,
dotted curves are $J$ = 2, lower solid curves are $J$ = 3.
For each $J$ the curve with stars is the L = J+1, the curve without
stars is L = J-1.}
\end{figure}

\begin{figure}
\caption[]{The phases $\delta_{\sl LJ}$ of the amplitudes of our fit
($J_{\sl max}$ = 3) versus p$_{\sl lab}$. The curves are labeled as in
Figure~3.}
\end{figure}

\begin{figure}
\caption[]{The total cross section (upper solid curve) for
$\overline{p}p \rightarrow \pi^- \pi^+$ as a function of p$_{\sl lab}$
for the solution from Table~II and III. Data points are from Ref.~\cite{LEAR}.
Contributions from $J$ = 0,1,2,3 are shown as respectively dashed,
dotdashed, dotted, and solid curves.
Again the curves are only meant to guide the eye. }
\end{figure}

\begin{figure}
\caption[]{Argand diagrams for the partial wave amplitudes
3P2 (solid with stars), 3F2 (solid), 3D3 (solid with stars),
3G3 (solid).
These amplitudes are taken from a $\chi^2$ fit with $J_{\sl max}$ = 3,
described in Table~II and III. The open circles indicate the lowest energy in
all the plots.}
\end{figure}

\begin{figure}
\caption[]{Fits of $\rm d\sigma/d\Omega$ and A$_{\sl on}$ at 783 MeV/c
for various sets
of J values as described in the text. Solid curves are for J = 0 -- 3,
dashed curves are for J = 0 -- 4, and dotted curves are for J = 1 -- 4.
The three curves are almost indistinguishable except for A$_{\sl on}$ at the
peak and in the extreme backward direction.}
\end{figure}

\begin{figure}
\caption[]{Predictions for A$_{\sl ls}$ and A$_{\sl ss}$ at 783
MeV/c. Solid line is for
regular solution (Table~II + III). Dashed curve is for
alternative solution (Table~IV + V).}
\end{figure}

\squeezetable
\begin{table}
\caption{Examples of $\chi^2$ per degree of freedom at each energy.}
\begin{tabular}{ddddddddd} 
p$_{\sl lab}$ (MeV/c) & $\chi^2$($J_{\sl max}$=2)& $\chi^2$($J_{\sl
max}$ = 3) &
$\chi^2$($J_{\sl max}$ = 4)\\
\tableline
360&  1.96&  1.77&  1.74\\
404&  1.38&  1.12&  1.12\\
467&  1.98&  1.31&  1.18\\
497&  3.04&  1.50&  1.45\\
523&  2.63&  1.45&  1.43\\
585&  1.96&  1.51&  1.57\\
679&  2.17&  1.50&  1.53\\
783&  2.50&  1.49&  1.47\\
886&  3.21&  1.23&  1.13\\
988&  4.39&  1.85&  1.55\\
\end{tabular}
\end{table}

\begin{table}
\caption{Energy dependence of parameters of $R_{\sl LJ}$ and
$\delta_{\sl LJ}$ for $J_{\sl max}$ = 3.}
\begin{tabular}{ddddddddd} 
p$_{\sl lab}$ (MeV/c) & $R_{10}$ & $\delta_{10}$ &
$R_{01}$ & $\delta_{01}$ &
$R_{21}$ & $\delta_{21}$
\\
\tableline
360&  1.60&  0 & 0.50& 120& 0.80& 0 \\
404&  1.44&  0 & 0.78& 140& 0.62& 20\\
467&  1.68&  0 & 0.88& 115& 0.64& 30\\
497&  1.60&  0 & 0.74& 115& 0.70& 15\\
523&  1.96&  0 & 0.90& 105& 0.78& 35\\
585&  1.62&  0 & 1.08& 125& 0.68& 75\\
679&  1.42&  0 & 1.22& 105& 0.88& 70\\
783&  1.22&  0 & 1.10& 105& 1.08& 95\\
886&  0.60&  0 & 0.80&  95& 1.32& 100\\
988&  0.14&  0 & 0.48& 175& 1.50& 185\\
\end{tabular}
\end{table}

\begin{table}
\caption{Energy dependence of parameters $R_{\sl LJ}$ and
$\delta_{\sl LJ}$ for $J_{\sl max}$ = 3.}
\begin{tabular}{ddddddddd} 
p$_{\sl lab}$ (MeV/c) & $R_{12}$ & $\delta_{12}$ &
$R_{32}$ & $\delta_{32}$ &
$R_{23}$ & $\delta_{23}$ &
$R_{43}$ & $\delta_{43}$
\\
\tableline
360&  0.28&   35& 0.34& -20& 0.10& -125& 0.06& 135\\
404&  0.16&   70& 0.48&   5& 0.22&  -55& 0.08& 125\\
467&  0.28&   95& 0.64&   0& 0.28&  -50& 0.20&  85\\
497&  0.46&   85& 0.48& -10& 0.32& -110& 0.36& 100\\
523&  0.34&  100& 0.38& -25& 0.26& -160& 0.40& 110\\
585&  0.34&  135& 0.74&   0& 0.28& -115& 0.34& 120\\
679&  0.44&  100& 0.92& -20& 0.32& -155& 0.42& 115\\
783&  0.58&  130& 1.02& -25& 0.52& -165& 0.48& 125\\
886&  0.56&  140& 1.14& -15& 0.72& -175& 0.60& 135\\
988&  0.58&  225& 1.06&  60& 0.66& -105& 0.60& 215\\
\end{tabular}
\end{table}

\begin{table}
\caption{Energy dependence of alternative parameter set
of $R_{\sl LJ}$ and $\delta_{\sl LJ}$.}
\begin{tabular}{ddddddddd} 
p$_{\sl lab}$ (MeV/c) & $R_{10}$ & $\delta_{10}$ &
$R_{01}$ & $\delta_{01}$ &
$R_{21}$ & $\delta_{21}$
\\
\tableline
360&  1.38&  0 & 0.92& 150& 0.44& -30\\
404&  1.12&  0 & 0.94& 140& 0.46& -10\\
467&  1.16&  0 & 1.06& 115& 0.34& -35\\
497&  0.82&  0 & 0.96& 115& 0.52& -40\\
523&  0.82&  0 & 1.24& 115& 0.64& -50\\
585&  0.84&  0 & 1.26& 150& 0.60& -70\\
679&  0.74&  0 & 1.30& 160& 0.72& -65\\
783&  0.66&  0 & 1.22& 220& 1.34&  -5\\
886&  1.42&  0 & 0.82& 235& 1.18&  35\\
988&  1.24&  0 & 0.66& 230& 1.04&  40\\
\end{tabular}
\end{table}

\begin{table}
\caption{Energy dependence of alternative parameter set
$R_{\sl LJ}$ and $\delta_{\sl LJ}$.}
\begin{tabular}{ddddddddd} 
p$_{\sl lab}$ (MeV/c) & $R_{12}$ & $\delta_{12}$ &
$R_{32}$ & $\delta_{32}$ &
$R_{23}$ & $\delta_{23}$ &
$R_{43}$ & $\delta_{43}$
\\
\tableline
360&  0.42&   90& 0.14&-105& 0.14&  100&  0.14&-210\\
404&  0.52&  105& 0.28& -40& 0.06&    0&  0.24&-180\\
467&  0.76&   95& 0.42& -25& 0.20&   25&  0.28&-195\\
497&  0.74&  100& 0.44& -10& 0.36&    0&  0.28&-160\\
523&  0.54&   80& 0.52&   5& 0.42&  -10&  0.26&-115\\
585&  0.64&  115& 0.68&  45& 0.32&   30&  0.30& -90\\
679&  0.70&  120& 0.92&  65& 0.40&   25&  0.36& -60\\
783&  0.50&  180& 0.88& 145& 0.46&   80&  0.56&  20\\
886&  0.54&  225& 0.92& 215& 0.50&  120&  0.94&  75\\
988&  0.38&  225& 0.92& 240& 0.46&  130&  1.06&  95\\
\end{tabular}
\end{table}

\end{document}